\documentclass{iau}
\usepackage{graphicx}

\title[Star formation at low and high redshift] 
{Theoretical considerations for star formation at low and high redshift}

\author[Bruce G. Elmegreen]   
{Bruce G. Elmegreen}

\affiliation{IBM T.J. Watson Research Center \\ 1101 Kitchawan Road, Yorktown Heights, NY, USA\\
email: {\tt bge@us.ibm.com}}

\pubyear{2015}
\volume{315}  
\jname{From Interstellar Clouds to Star-Forming Galaxies: Universal Processes?}
\editors{Pascale Jablonka, Philippe Andr\'e \& Floris van der Tak, eds.}
\begin{document}

\maketitle

\begin{abstract}
Star formation processes in strongly self-gravitating cloud cores should be similar at all redshifts, forming
single or multiple stars with a range of masses determined by local magneto-hydrodynamics and gravity. The
formation processes for these cores, however, as well as their structures, temperatures, Mach numbers, etc., and
the boundedness and mass distribution functions of the resulting stars, should depend on environment, as should
the characteristic mass, density, and column density at which cloud self-gravity dominates other forces. Because
the environments for high and low redshift star formation differ significantly, we expect the resulting gas to
stellar conversion details to differ also. At high redshift, the universe is denser and more gas-rich, so the
active parts of galaxies are denser and more gas rich too, leading to slightly shorter gas consumption
timescales, higher cloud pressures, and denser, more massive, bound stellar clusters at the high mass end. With
shorter consumption times corresponding to higher relative cosmic accretion rates, and with the resulting higher
star formation rates and their higher feedback powers, the ISM has greater turbulent speeds relative to the
rotation speeds, thicker gas disks, and larger cloud and star complex sizes at the characteristic Jeans length.
The result is a more chaotic appearance at high redshift, bridging the morphology gap between today's quiescent
spirals and today's major-mergers, with neither spiral nor major-merger processes actually in play at that time.
The result is also a thick disk at early times, and after in-plane accretion from relatively large clump torques,
a classical bulge. Today's disks are thinner, and torque-driven accretion is slower outside of inner barred
regions. This paper reviews the basic processes involved with star formation in order to illustrate its evolution
over time and environment. 
\end{abstract}

\firstsection 
\section{Introduction}

Star formation (SF) in young L* galaxies looks very different from local SF. At a redshift of around 2, galaxies
have a high gas fraction close to 50\% (\cite[Tacconi et al. 2013]{tacconi13}; \cite[Duncan et al.
2014]{duncan14}), a high ratio of gas velocity dispersion to rotation speed of 30\% or more
(\cite[F\"orster-Schreiber et al. 2009]{forster09}; \cite[Kassin et al. 2012]{kassin12}), a high ratio of
interstellar Jeans Length to galaxy radius of 30\%, which follows from the high dispersion and gives big clumps
of SF and a thick star-forming disk, a theoretically predicted high accretion rate (e.g., \cite[Dekel et al.
2013]{dekel13}), which produces a high SF rate, and strong outflows from the high SF rate (\cite[Newman et al.
2012]{newman12}). High redshift galaxies also seem to produce more massive clusters than today's galaxies (see
below), perhaps because of their larger ISM pressures.

SF today looks very different. L* galaxies have a low gas fraction, $\sim5$\%, the ratio of dispersion to
rotation speed is low, $\sim10$\%, producing relatively small star-forming clumps in a thin disk, the accretion
rate is low, which contributes to the low SF rate, outflows are relatively weak, and the maximum cluster mass is
lower.

Even with these differences, the SF processes themselves appear to be uniform for young and old galaxies. This
uniformity is manifest in several ways, as discussed in the next section.

\section{Uniformity in star formation processes}

We enumerate here the many ways in which star formation processes appear to be similar over a wide range of
redshifts.

1. Most star-forming galaxies lie on a main sequence in a plot of SF rate versus galaxy mass. This main sequence
has a constant slope with redshift, up to $z\sim2.5$, with a change only in the absolute SF rate for a given
galaxy mass, which increases at higher redshift (\cite[Whitaker et al. 2012]{whitaker12}). The constant slope
suggests that the SF process is uniform on a galactic scale from galaxy to galaxy. More massive galaxies have
more of the same processes of star formation.

The upward shift in the SF rate per unit stellar mass, or the specific SF rate, sSFR, satisfies approximately
sSFR$\propto(1+z)^{2}$ for $\log M_{\rm star}\sim9.7\pm0.3$ from SED and nebular line emission (\cite[Duncan et
al. 2014]{duncan14}). What drives this shift is probably not a change in process, but rather a change in the
available fuel, i.e., the relative gas mass (\cite[Tacconi et al. 2013]{tacconi13}; \cite[Duncan et al.
2014]{duncan14}). This increased gas mass follows from an increased gas accretion rate, according to cosmological
models (\cite[Dekel et al. 2013]{dekel13}).

2. The Kennicutt-Schmidt relation is nearly invariant with redshift for star-forming galaxies (\cite[Tacconi et
al. 2013]{tacconi13}, see also \cite[Genzel et al. 2010]{genzel10}; \cite[Daddi et al. 2010]{daddi10};
\cite[Saintonge et al. 2012]{saintonge12}; \cite[Magdis et al. 2012]{magdis12}). This implies that the gas
depletion time is about constant with redshift, varying only by a factor of $\sim2$ up to $z\sim4$, considering
observations of gas mass from either molecular or dust emissions (\cite[Genzel et al. 2015]{genzel15};
\cite[B\'ethermin et al. 2015]{bethermin15}). This is a much slower variation than expected if the SF rate scales
with the average internal rotation rate of a galaxy.

3. Fluctuations in the SF rate around the SF main sequence are small in the disk, corresponding to a factor of
$\sim2$ at both low (\cite[Abramson et al. 2014]{abramson14}; \cite[Guo et al. 2015]{guo15}; \cite[Casado et al.
2015]{casado15}) and intermediate (\cite[Salmi et al. 2014]{salmi14}) redshifts. According to \cite[Casado et al.
(2015)]{casado15} who studied 82,500 SDSS galaxies, H$\alpha$ and u-v colors show an aging sequence with internal
evolution for field galaxies driven by accretion, and then a rapid quenching of SF in dense environments.

4. The normalized mass-metallicity relation is uniform over redshift. \cite[Zahid et al. (2014)]{zahid14} showed
that the redshift dependent mass-metallicity relation for galaxies can be scaled so that different redshifts lie
on a common relationship. The scaling is set by the stellar mass where the gas and star masses are equal in a
galaxy, and by a peak metallicity where the metal production rate by stars minus the loss rate in winds equals
the metal lock-up rate in stars. With this scaling, the relationship between oxygen metallicity and the
gas-to-star ratio is constant. Thus the apparent redshift evolution of the mass-metallicity relation is from the
varying gas-to-star ratio, independent of SF processes.

5. There is uniformity in feedback over galaxy mass and cosmic time in the form of a nearly constant M* at the
break of the galaxy mass function (\cite[Song et al. 2015]{song15}), which is probably determined by feedback
(similar results are in \cite[Duncan et al. 2014]{duncan14}; \cite[Grazian et al. 2015]{grazian15}).

6. There is uniformity over a wide redshift range in the dichotomy between SF galaxies and ULIRGs/mergers, which
have higher efficiencies corresponding to lower depletion times (\cite[Tacconi et al. 2013]{tacconi13}; see also
\cite[Genzel et al. 2010]{genzel10}; \cite[Daddi et al. 2010]{daddi10}; \cite[Saintonge et al.
2012]{saintonge12}; \cite[Magdis et al. 2012]{magdis12}). This dichotomy was also observed by \cite[Silverman et
al. (2015)]{silverman15} using ALMA and IRAM/PdBI CO observations of z=1.6 starburst galaxies and a molecular
mass from the virial theorem. It was observed by \cite[B\'ethermin et al. (2015)]{bethermin15} using molecular
mass from dust observations and by \cite[Sargent et al. (2014)]{sargent14}, who calibrated molecular mass using
an H$_2$/CO conversion factor that depends on metallicity (which depends on SF rate and stellar mass).

The usual explanation for this dichotomy is that the star formation time scale in gas of a given mass depends on
the average density. In ULIRGs and mergers, torques shrink the gas disk and increase the density. Shrinkage by
1/4 for example, from $R_{25}$  to the disk scale length, causes the gas mass surface density to increase by a
factor of 16 and the pressure to increase by a factor of 256 (scaling as the square of the surface density). Thus
an increase in density by a factor of 100 is not unreasonable, and this would decrease the free fall time for a
given column density by a factor of 10. In addition, mergers create large shocked regions where the ISM from the
two galaxies crashes together, and this makes high densities too.  \cite[Teyssier et al. (2010)]{teyssier10}
showed this effect in numerical simulations: mergers make dense gas with a short consumption time, placing them
higher on the KS relation for a given gas column density.

\section{What are the unifying processes of star formation?}

What is the unifying physics in the star formation main sequence? \cite[Krumholz, Dekel \& McKee
(2012)]{krumholz12} suggest it is feedback-controlled disk stability that makes the Toomre Q always about 1 or 2.
If we write the star formation rate per unit gas mass as $ \Sigma_{\rm SFR} = f(H_2) \epsilon_{\rm ff}
\Sigma_{\rm gas} / \tau_{\rm ff}$ for molecular fraction $f(H_2)$, constant efficiency per unit free fall time
$\epsilon_{\rm ff}$, gas mass surface density $\Sigma_{\rm gas}$, and free fall time $t_{\rm ff}$, and take for
$t_{\rm ff}$ the minimum value of the free fall time in giant molecular clouds and the average free fall time for
a $Q\sim1$ disk, then the Kennicutt-Schmidt relation is a uniform linear function of the areal SF rate versus
$\Sigma_{\rm gas}/t_{\rm ff}$.  The Toomre free fall time comes from the average midplane density
$\rho\sim\Sigma_{\rm gas}/2H$ for scale height $H\sim\sigma^2/(\pi G \Sigma_{\rm gas})$. When $Q\sim1$ by
self-regulation, then $\sigma\propto\Sigma_{\rm gas}$ at constant $\kappa$ because $Q=\sigma\kappa/(\pi
G\Sigma_{\rm gas})$ for velocity dispersion $\sigma$ and epicyclic frequency $\kappa$.  When $\sigma$ is
proportional to $\Sigma_{\rm gas}$, then $H$ is also proportional to $\Sigma_{\rm gas}$ and the midplane density
is about constant. Higher areal SF rates then correspond to thicker disks at constant $\kappa$.

The self-regulation of $Q$ in disks is commonly seen in numerical simulations when spirals are active. Low $Q$
makes spirals which stir the stars and gas and increase $Q$. The application of $Q$ to star formation is more
indirect and contains assumptions which may not be correct in some cases. Constant $Q$ reflects a balance between
self-gravity and centrifugal force in a radially contracting perturbation that was spun up by the Coriolis force.
$Q=1$ when these two forces balance on the scale of the fastest growing perturbation, which is twice the
threshold Jeans length.

There are problems with applications of an instability parameter like $Q$ to gas and SF. First, the gas is easily
torqued by magnetic fields and viscosity, which removes the angular momentum and reduces spin-up. Second, gas
turbulent energy always dissipates, removing the Jeans length as a minimum scale (\cite[Elmegreen
2011]{elmegreen11}). Third, the instability condition was derived for rings (spirals have an additional parameter
for instabilities in the azimuthal direction) and is applicable on a scale equal to $2\pi G\Sigma/\kappa^2$,
which is comparable to a spiral arm spacing. It is 2-dimensional and not directly connected with 3D collapse in
midplane gas. There are also thickness effects that should be considered explicitly in the stability condition
even for two-dimensional instabilities (e.g., \cite[Romeo \& Wiegert 2011]{romeo11}).

In a study of 20 dwarf Irregular galaxies, \cite[Elmegreen \& Hunter (2015)]{elmegreen15} used the observed
rotation speed to determine $\kappa$, and the observed values of $\Sigma_{\rm gas}$, $\sigma_{\rm gas}$ and
$\Sigma_{\rm star}$, along with a scaling of $\sigma_{\rm star}$ with absolute galaxy magnitude, to determine $Q$
for the gas and stars in radial annuli. The gas $Q$ averaged about 4 and the stellar $Q$ was about 10 or larger.
These values are so high that even the combined $Q$ for a 2-fluid instability (\cite[Romeo \& Wiegert
2011]{romeo11}) would be high. Yet SF is pervasive and normal-looking without $Q \sim 1-2$ as in spirals.
Evidently, SF and Q do not self-regulate to make $Q \sim 1-2$ everywhere, and moderately high $Q$ does not stop
SF, even though it apparently stops spiral arms from forming (as in dwarfs).

The question about the relevance of $Q$ to star formation comes up again for the central regions of star-forming
galaxies where a high $Q$ from the bulge is suggested to cause ``morphological quenching'' (\cite[Martig et al.
2009]{martig09}; \cite[Genzel et al. 2014]{genzel14}). This quenching in a bulge is a clear effect in
observations and simulations, but the cause of it could be something other than a high $Q$. For example, it could
be an effect of a high critical tidal density for clouds to be self-bound, or a lack of spirals inside the inner
Lindblad resonance. The tidal density for a co-rotating cloud is $\rho_{\rm tidal}=(3/2\pi G)\Omega^2$ for
rotation rate $\Omega$. That is a three-dimensional threshold for ambient gas to turn into self-gravitating
clouds, independent of spiral arms. This should be high in bulge regions because $\Omega$ is high, regardless of
$Q$. A condition for bulge quenching would then be $\rho<<\rho_{\rm tidal}$ for ambient density $\rho$.

Analytical models assume that the SF rate is the collapse rate at all densities above some critical value in a
density distribution function that is log-normal from turbulence. \cite[Federrath \& Klessen (2012)]{federrath12}
compared several analytical models of SF to simulations and found that they agreed with each other to within a
factor of 2. The models also agreed reasonably well with observations.  This agreement suggests that various
details in the models, such as the dependence of the critical density on model parameters, do not matter much for
the overall SF rate.  Federrath et al. also note that the power law form of the density pdf, which develops after
the gas begins to be strongly self-gravitating, does not significantly affect the predicted rates.  Power law
forms for this pdf now seem preferred (\cite[Lombardi et al. 2015]{lombardi15}). \cite[Federrath \& Klessen
(2012)]{federrath12} suggest that the log-normal density pdf is still appropriate for the initial conditions in
the models.

The dual-track Kennicutt-Schmidt relation with one track for normal star forming galaxies and another for
starbursts (see above), could have a contribution from metallicity changes. \cite[Tan et al. (2013)]{tan13} note
that relatively low CO emission at a given SF rate for high redshifts ($z>3$), which is another way of expressing
the second track from starbursts, might be from low metallicity instead of high density.

Local observations show that CO emission drops significantly at low-metallicity for a given molecular mass or
star formation rate (\cite[Bolatto et al. 2013]{bolatto13}; \cite[Leroy et al. 2011]{leroy11}). CO emission in
the local group galaxy WLM, which has a metallicity of 13\% solar, is very weak compared to the mass of molecular
hydrogen inferred from dust and HI emission (\cite[Elmegreen et al. 2013]{elmegreen13}). ALMA observations of the
individual CO clouds also show them to be very tiny compared to the extent of the H$_2$ (\cite[Rubio et al.
2015]{rubio15}). Nevertheless, the clouds are like small CO clouds in the solar neighborhood: they have about the
same pressure, density and column density, and they satisfy the usual size-linewidth relation for other dwarf
galaxies, although at the low mass end. Morphologically, the CO in WLM is highly confined to the core regions of
much larger H$_2$ clouds. This is unlike CO clouds in the Milky Way which share more of the H$_2$ volume. The
conversion from CO emission to H$_2$ mass is $\sim30$ times higher in WLM than in the Milky Way because of this
confinement.

The centralized location for CO seems to be necessary for low-pressure galaxies like WLM because the CO needs a
certain column of dust and H$_2$ to shield it from background radiation, and it needs a high density for
excitation which cannot come from the ambient ISM pressure alone. What gives it this high density is the large
weight of the overlying H$_2$ and HI layers, which both shield it and compress it (\cite[Rubio et al.
2015]{rubio15}).

Galaxies at high redshift might show the same centralized structure for CO if the metallicity is low. The change
would be in the direction of greater SF rate per CO luminosity, which is what also happens in ULIRGs. The reason
for this change would be different in the two cases because in ULIRGS, the CO gas should be denser than usual,
causing faster SF per unit CO mass, and at low metallicity, the CO should be more sparse than usual, causing
normal SF rates but with less CO.

\section{Massive Clusters at high redshift}

A possible implication of the small CO cores in local metal-poor galaxies is that only small star clusters might
form. This would be the case if SF happened only in CO-emitting regions. If SF also occurs in cold, shielded
regions without CO (\cite[Glover \& Clark 2012]{glover12}; \cite[Krumholz, Leroy \& McKee 2011]{krumholz11}) then
clusters more massive than the CO cloud could form. In either case, it seems possible that low-mass,
low-metallicity galaxies left on their own will not be able to make bound star clusters as massive as the metal
poor globular clusters currently seen in galactic halos. These clusters probably formed in dwarf galaxies
(\cite[Searle \& Zinn 1978]{searle78}; \cite[Elmegreen, Malhotra, \& Rhoads 2012]{elmegreen12}) when the
metallicity was as low as we see it today in the clusters. The clusters could also have been much more massive
when they formed, considering the evaporation of stars in the intervening time (\cite[McLaughlin \& Fall
2008]{mf08}). Thus there could be a problem in understanding how low-mass and low-metallicity galaxies formed
massive bound clusters. A possible solution is that the clusters formed during mergers between such galaxies,
when a large mass of interstellar material would have been at high pressure (\cite[Bekki 2008]{bekki08};
\cite[Rubio et al. 2015]{rubio15}).

Cluster formation at high redshift seems to have made more massive bound clusters than are being produced today.
\cite[Jordan et al. (2007)]{jordan07} found that the Schechter cutoff mass for globular cluster systems decreases
with galaxy mass but in all cases, this cutoff is about 10 times higher than the cutoff mass seen in local
galaxies of the same mass, even including the Antenna galaxy merger.  In \cite[Jordan et al. (2007)]{jordan07},
the cutoff mass for globular clusters associated with galaxies having an absolute blue magnitude of $-21.6$ is
around $2\times10^6\;M_\odot$, whereas in M51 and the Antenna, which have the same magnitude, it is
$10^5\;M_\odot$ and $4\times10^5\;M_\odot$, respectively (\cite[Gieles et al. 2006]{gieles06}).

\cite[Zaritsky et al. (2015)]{zaritsky15} identified globular clusters in the Spitzer Survey of Stellar Structure
in Galaxies (S$^4$G). They calculated the number of globular clusters per $10^9\;M_\odot$ of galaxy mass. For
dwarfs, approximately 2\% of today's galaxy mass is in globular clusters. This means that at 10\% of the
universe's age, when the clusters formed, and for ten times the current cluster mass in the initial state (as
constrained by the observation of dual populations -- \cite[D'Ercole et al. 2012]{dercole12}), the total globular
cluster mass was comparable to the galaxy mass at the time of globular cluster formation. Then the dwarf galaxy
would have been dominated by giant clusters. The same high fraction is observed for WLM (\cite[Elmegreen,
Malhotra, \& Rhoads 2012]{elmegreen12}) and Fornax dSph (\cite[Larsen et al. 2012]{larsen12}).

Cluster formation thus appears to have been more efficient at high redshift than it is today. If the formation
efficiency of bound clusters is higher at higher pressure and higher star formation rate (\cite[Larsen
2009]{larsen09}; \cite[Kruijssen 2012]{kruijssen12}; \cite[Adamo et al. 2015]{adamo15}) then perhaps the greater
cutoff mass of clusters at high redshift also followed simply from the higher gas fraction, like the other
changes with redshift discussed above.

\section{Conclusions}

SF processes look the same for a wide range of cosmic time. The SF Main Sequence and constant M* at the cutoff in
the galaxy mass function suggest a uniformity of SF among different galaxies. The small dispersions in the SF
main sequence, the Kennicutt-Schmidt relation, and the mass-metallicity relation all suggest a uniformity of SF
within galaxies.  There also appears to have been the same distinction between ULIRGs and normal star-forming
galaxies for a wide range of cosmic time, leading to higher SF efficiencies and shorter gas consumption times in
mergers, presumably from the higher densities in shocks and inflows in these systems.

Stars form in dwarf irregular galaxies even when the Toomre Q parameter is high. Thus the validity of a $Q$
threshold for SF in dwarfs is questionable, and, by analogy, also questionable in the outer parts of spirals,
which have similar properties. The applicability of a $Q$ threshold to quenching of SF in the bulge regions of
spirals might also be questioned. The utility of $Q$ may be limited to spiral-scale instabilities in thin disks.

The microscopic SF mechanism is probably universal: turbulent and directed compressions of cold gas in
self-gravitating clouds, followed by local collapse. Low metallicity at high redshift should change the
large-scale correlations by changing the conversion factor between CO emission and cold gas mass, but the
microscopic processes of SF could still be about the same.

Even with these similarities, the first clusters that formed appear to extend to much
higher masses than the largest clusters forming today. Perhaps this reflects some
differences in physical processes, such as less feedback at lower metallicities or more
prevalent galaxy collisions in early times, or perhaps it is another indication, similar to
what is starting to be observed locally, that cluster formation efficiency increases with
pressure. Higher pressures are expected at high redshift because of the greater gas
fractions and densities of disks.

\end{document}